# Learning Electromagnetism through a playful fair-game project

Arturo Pazmino[1*], Luis Pabón[1], Esther Desiree Gutiérrez M.[1], Erick Lamilla[1], and Eduardo Montero[1]

[1]Escuela Superior Politécnica del Litoral, ESPOL, Departamento de Física, Camus Gustavo Galindo km 30.5 Vía Perimetral, P.O. Box 09-01-5863, Guayaquil, Ecuador

*Corresponding author: apazmino@espol.edu.ec

Project/problem-based learning (PBL), as an active methodology, improves significantly the learning process, making students take an active role in the construction of their own knowledge, and at the same time, develop soft and social skills that are critical in the success of their student career and professional field. In this work, an entertaining game project based on an introductory undergraduate physics course is presented, in which students build an experimental prototype based on a traditional fair-game, "High Striker" game. To fulfill the project's requirements, students need to use mainly electromagnetism concepts such as Faraday's law, Lenz's law, induced electromotive force; and classical mechanics physics concepts such as energy conservation and collisions.

PBL is considered one of the active learning methodologies more significant in the last twenty years, due to the active role that students have in their learning process, promoting academic motivation inside the classroom[1,2]. PBL seeks a group interaction among students through the analysis and solution of a real-life project or challenging problem, developing critical thinking, decision-making [3,4,5,6], group collaboration and responsibility, developing a commitment attitude[7], and the teacher taking a guide role in this learning process[8].

## Fair-game project: Test your strength (High striker)

This project is based on the traditional county fair-game, "High Striker". The idea is to hit a target with a mallet to raise a spinning top and hit a bell to win a prize, thus you need to show a lot of strength (or probably know physics tricks) to hit the bell. This project is based on that game and the main goal is to design and build an electromagnetic propulsion system actuated by a mechanical interaction that allows a sphere to move along a track, showing concepts of electromagnetic induction and energy conservation. For a better follow up and feedback, the implementation of this project is required to be modular, each component can be designed, built, and implemented separately. Then, the project is divided into three modules plus the track.

*The first module is the force interaction module*, consisting of a hammer and a target, once the player hits the target, a certain light emitter diodes (leds) lid by means of a mechanical-electrical transducer. The lit diodes will depend on the amount of force given in the target, different colors can be used, such that red diode means low amount of force, yellow diode represents medium force and green diode, high force. The range for a low, medium, and high force will depend on each prototype, students will define these ranges according to their tracks.

*The second module is the shooting module*, it consists in the construction of a homemade coil that launches a latch, using concepts of electromagnetic induction, to hit the sphere and make it move along the track. This coil should be activated once a current passes through it in a certain time interval. The development of this module involves the knowledge of electromagnetism concepts such as Ampere's, Biot-Savart's and Faraday's law and their applications, for the designing parameters to build a solenoid and how they affect the theoretical calculations of the magnetic field.

*The third one is the signal adaptation module*, it should be designed and implemented such that its entry signal is coming from the force interaction module, and the output signal allows the current to circulate around the coil in the shooting module. A programable interpreter needs to be used to manipulate the analog input signal and produce an output signal, this interpreter is commonly an Arduino.

The track structure is also important when designing the modules, it must have a minimum length, then the steeper the track the more energy is needed for the sphere to reach the top, and more adjustments need to be done in the circuit to get that energy. For this part, energy calculations in each relevant point of the prototype are needed, considering translational and rotational kinetics energies, potential energy, and including energy losses, even if they are minimal, such as during the collision between the latch and the sphere, due to small deformation, sound and heat, and possible friction force during the movement of the latch and along the track. Another consideration is about the sphere, it cannot be metallic because it will be attracted by the induced electromagnet in the shooting module.

## Results

The project was worked in an introductory electromagnetism physics course of an undergraduate level. Students were divided into teams, between six to seven students per team. Students start the development of the project by interpreting relevant information such as limitations, fundamental parts of the prototype, and basic classical mechanics and classical electromagnetism concepts.

Briefly, the game starts by hitting a target (force interaction module), which is a sensor, and this signal is interpreted by an Arduino board (signal adaptation module) which will turn on certain led with a specific color that visually indicates the amount of force given to the target. The Arduino also emits a signal that activates the homemade coil and, by magnetic induction, move the latch that will shoot the sphere (shooting module) along the track.

Figure 1 shows the schematic of a typical solution simulation of the project. The force interaction module is made of a transducer that feeds the Arduino. Some solutions used a piezoelectric with a non-inverting amplifier based on an operational amplifier (OpAmp), to increase the value of the input signal to the Arduino (741 OpAmp[9] is the common integrated circuit for this purpose). This increment is usually necessary since students tend to hit too hard that might break the sensor, then a flex module adapted to a plastic sheet is used to cushion the impact. The flex module increases its electrical resistance with this mechanical stress placed on it, causing a low input signal to the input pin of the Arduino. A button is adapted to simulate a hit while it is pressed, HIT_SIMULATOR (only for the simulation).

In the signal adaptation module, the Arduino interprets the analog input and sends a signal to the corresponding digital output pin with the red, yellow, or green led. There is a standby led (blue) to announce the expectation of a hit, which turns off once the Arduino detects the input signal coming from the force interaction module. When the amount of force, from the hit, falls into one of the three categories, low, medium, and high, the output signal triggers the shooting module.

The shooting module consists of a set of relays, resistors, and a voltage source (typically 12V and 2 A), all of them are connected to a solenoid that will active the latch, the solenoid is represented in the schematic by the Motor. If the hit falls in the low range force, the first relay is activated (KRELAY_DISP_LOW), the solenoid sees the voltage source and a high resistance pot (RPOT_SHOO_LOW) connected in series. If it falls in the medium range force, the second relay is activated (KRELAY_DISP_MIDDLE), the solenoid sees the voltage source and a low resistance pot (RPOT_SHOO_MID) connected in series. And finally, for the high range force, the third relay is activated (KRELAY_DISP_HIGH) and the solenoid is connected directly to the voltage source. The current through the solenoid lags behind the change in voltage across its terminals, then its self-induction pulls the latch that will hit the sphere. Another way to do it is that each relay charges a specific bank of capacitors, and the current produced by the discharging capacitors feeds the coil to generate an impulse of current and therefore a self-induction.

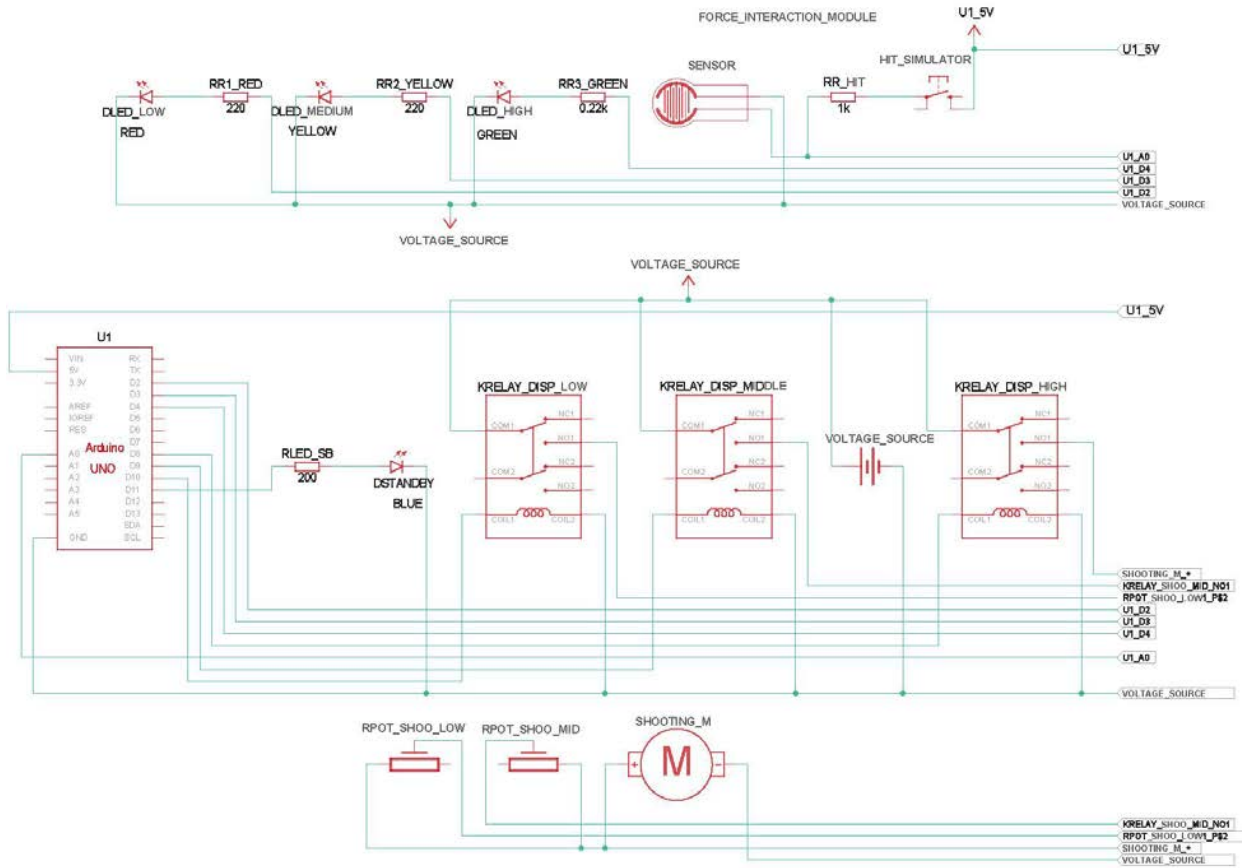

*Figure 1: Schematic diagram of a typical solution of the test your strength project*

The most common elements used in each module by the prototypes (shown in the schematic above) are listed in Table 1.

| Item | Quantity | Element | Module |
|---|---|---|---|
| SENSOR | 1 | Force Sensor | Force interaction |
| HIT_SIMULATOR | 1 | Button (only used in the simulation) | Force interaction |
| RR_HIT | 1 | 1 kΩ Resistor | Force interaction |
|  |  |  |  |
| U1 | 1 | Arduino Uno R3 | Signal adaptation |
| RR1_RED | 1 | 220 Ω Resistor | Signal adaptation |
| RR2_YELLOW | 1 | 220 Ω Resistor | Signal adaptation |
| RR3_GREEN | 1 | 0.22 kΩ Resistor | Signal adaptation |
| DLED_LOW | 1 | Red LED | Signal adaptation |
| DLED_MEDIUM | 1 | Yellow LED | Signal adaptation |
| DLED_HIGH | 1 | Green LED | Signal adaptation |
| DStandby | 1 | Blue LED | Signal adaptation |
| RLed_SB | 1 | 200 Ω Resistor | |
| | | | |
| KRELAY_DISP_LOW | 1 | Relay DPTP | Shooting |
| KRELAY_DISP_MIDDLE | 1 | Relay DPTP | |
| KRELAY_DISP_HIGH | 1 | Relay DPTP | |
| SHOOTING_M | 1 | Motor (represents the homemade coil in the simulation) | |
| RPOT_SHOO_LOW | 1 | 500 Ω Potentiometer | |
| RPOT_SHOO_MID | 1 | 500 Ω Potentiometer | |
| | | | |
| VOLTAGE_SOURCE | 1 | Voltage supply, 12V and 2 A | Power Supply |

*Table 1: List of components and supplies used by the project*

The Arduino code that interprets the input signal from the force interaction module and generates an output signal to the shooting module is provided in the supplementary material, Section A.

# Conclusions

This work shows an entertainment project based on the traditional county fair-game in a town, "High strike". The develop solution of this project is based mostly on classical electromagnetism concepts, and also includes some concepts of classical mechanics physics. Students were able to build their own knowledge on electromagnetism through an active learning process methodology such as project-based learning, and they also reinforced knowledge learned in a previous physics course.

The performance of the teams was remarkable. The final design of their prototypes was eye-catching, and most of them, during the final oral presentation, seemed to have a very good understanding of the physics behind the project. Pictures of several prototypes of the Test your strength playful game project is shown in figure S1 in the supplementary material, section B. Feedback from some students, about the project, is also shown in Section B of the supplementary material.